\documentclass[conference]{IEEEtran}

\usepackage{graphicx,mathptmx,amsmath,amsfonts,amsthm,color,comment}
\usepackage{enumitem}
\usepackage{enumerate}
\usepackage{algorithmic}
\usepackage{pgfplots}
\usepackage{pgf}
\usepackage{tikz}

\def\ben{\begin{eqnarray*}}
\def\een{\end{eqnarray*}}

\def  \codebook{C}

\def  \H5G{H_\text{CA-Polar}}

\def  \C{\mathbb {C}}

\def  \R{\mathbb {R}}

\def\soft{ \Phi }

\def\R{\mathbb{R}}

\def\mod{\text{mod}}

\begin{document}


\title{Soft decoding without soft demapping with ORBGRAND}

\author{

\IEEEauthorblockN{Wei An}
\IEEEauthorblockA{\textit{RLE} \\
\textit{MIT}\\
Cambridge, MA, USA \\
wei\_an@mit.edu}
\and
\IEEEauthorblockN{Muriel M\'edard}
\IEEEauthorblockA{\textit{RLE} \\
\textit{MIT}\\
Cambridge, MA, USA \\
medard@mit.edu}
\and
\IEEEauthorblockN{Ken R. Duffy}
\IEEEauthorblockA{\textit{Hamilton Institute} \\
\textit{Maynooth University}\\
Ireland \\
ken.duffy@mu.ie}
}

\maketitle

\begin{abstract}
For spectral efficiency, higher order modulation symbols confer information on more 
than one bit. As soft detection forward error correction decoders assume the availability of information at binary granularity, however, soft demappers are required
to compute per-bit reliabilities from complex-valued signals. Here we show that the recently introduced universal soft detection
decoder ORBGRAND can be adapted to work with symbol-level soft information, obviating the need for energy expensive soft demapping.
We establish that doing so reduces complexity while retaining the error correction performance achieved with the optimal demapper.
\end{abstract}


\section{Introduction}

To enable a receiver to detect or correct errors, prior to transmission each collection of $k$ information bits is coded to a $n>k$ bit code-word $c^n=(c_1,\ldots,c_n)\in\{0,1\}^n$. For example, CRC-Assisted Polar (CA-Polar) codes will be used for
all control channel communications in 5G New Radio, while Low Density Parity Check codes will be used for all data transmissions \cite{3gpp38212}. For spectral efficiency, most communication systems employ high-order modulation, such as Quadrature Amplitude Modulation (QAM), where each transmitted symbol communicates multiple bits of information \cite{Proakis}.
If a modulation scheme is employed with a complex constellation of size $|\chi|=2^{m_s}$, the $n$ coded bits are translated into $n_s=n/m_s$ symbols by sequentially mapping each collection of $m_s$ bits to the corresponding higher order symbol, resulting in the transmission of the higher order sequence $\mod(c^n) = x^{n_s} = (x_1,\ldots,x_{n_s}) \in \chi^{n_s}$.

Transmissions are impacted by noise and channel effects resulting in the received signal sequence being perturbed. If these effects are written as $N^{n_s}=(N_1,\ldots,N_{n_s})\in \C^{n_s}$, then the complex received vector can be expressed as $Y^{n_s} = (Y_1,\ldots,Y_{n_s}) = x^{n_s}+N^{n_s}$. As essentially all soft detection forward error correction decoders require binary input and per-bit reliability information \cite{shu2004}, the existing paradigm is to process the complex received signals, $Y^{n_s}$, and so evaluate a sequence of per-bit reliability metrics, $\gamma^n \in \R^n$, via a soft demapper. 


Extracting bit-level soft information from higher order signals is a computational complex and energy expensive process, even in the simplest setting where noise is additive and a symbol level interleaver is employed so that the noise effects can be assumed to be independent and identically distributed at the level of symbols. To grasp why, consider the conventional log-domain Maximum {\it a posteriori} (Log-MAP) demapping algorithm \cite{Erfanian1994}. With $\chi_l^{(j)}$ representing the subset of constellation symbols where the $l$-th bit in their binary representation, $l\in\{1,\ldots,m_s\}$, takes the value $j\in\{0,1\}$ and $f_{Y|X}(Y|x)$ being the conditional probability of observing the signal $Y\in\C$ given $x\in\C$ was transmitted, then the log-likelihood ratio (LLR) for bit position $i$ in the original code-word, whose information is contained in the symbol $x_{\lceil i/m_s\rceil}$, is determined by the following demapping:
\begin{align} 
    &\gamma_i = \log {\frac{f_{Y|X}\left(Y_{\lceil i/m_s \rceil}|x\in \chi_{i-(\lceil i/m_s\rceil-1) m_s}^{(1)} \right)}{f_{Y|X}\left(Y_{\lceil i/m_s \rceil}|x\in \chi_{i-(\lceil i/m_s\rceil-1) m_s}^{(0)}\right)}} \nonumber \\
    &= \log \left( \frac{\sum_{x\in \chi_{i-(\lceil i/m_s\rceil-1) m_s}^{(1)}} \limits f_{Y|X}(Y_{\lceil i/m_s \rceil}|x) }{ \sum_{x\in \chi_{i-(\lceil i/m_s\rceil-1) m_s}^{(0)}} \limits f_{Y|X}(Y_{\lceil i/m_s \rceil}|x) } \right). \label{eq:demap}
\end{align}
The evaluation of eq. \eqref{eq:demap} requires order $|\chi|$ operations per transmitted bit and involves the computation of a logarithm as well as exponentials, assuming the usual additive Gaussian noise model, resulting in determination of a LLR for each the $n$ demodulated bits. As demapping is an essential element of soft detection error correction decoding systems, substantial ongoing research seeks to identify more energy-efficient approximate LLR calculations, e.g. \cite{Robertson1995,Tosato2002, Akay2004, Chang2006, Wang2014, Mao2016}.

While traditional soft detection decoding algorithms for binary codes can only work with binary soft information, here we establish that a soft detection
variant of Guessing Random Additive Noise Decoding (GRAND) is capable of operating directly with soft information at the level of symbols, obviating the need for costly soft demapping. 
Moreover, in the process of demapping, information about which bits are coupled to each other in a symbol is irretrievably lost, and making use of that information is shown to enable reduced complexity.

\section{Guessing Random Additive Noise Decoding}
GRAND is a recently established approach that can be used to efficiently decode any moderate redundancy code. Originally considered for hard decision demodulation systems \cite{duffy19GRAND,An22}, soft detection variants that assume the availability of per-bit soft information from demappers have since been developed \cite{solomon20SGRAND,duffy2021ordered,Duffy22,duffy2022ordered}. Such is GRAND's practical promise that implementations
have already been published for both the hard detection \cite{abbas2020, abbas2021high-MO,Riaz21,zhan2021noise,Riaz22} and soft detection settings \cite{condo2021high,condo2022fixed,abbas22}.

GRAND's operation is readily understood, and pseudo-code for it can be found in Fig. \ref{alg:pseudo-code}. As a descriptor of a code, GRAND solely requires a function that, given a string, reports whether it is in the codebook. GRAND algorithms can be used to decode standard and non-standard linear codes, such as Cyclic Redundancy Check (CRC) and Random Linear Codes (RLCs) \cite{an21, Papadopoulou21}, as well as entirely new non-linear codes constructed from cryptographic functions \cite{cohen22}. Based on the statistical description of the channel or soft information, GRAND algorithms generate discrete noise-effect patterns in order from most likely to least likely. By  sequentially subtracting these patterns from the received signal and querying if what remains is an element of the codebook, the first identified codeword is an optimally accurate Maximum Likelihood (ML) decoding.

\begin{figure}
\hrule
\noindent
\begin{algorithmic}
\STATE {\bf Inputs}: Code-book membership function $\codebook:\{0,1\}^n\mapsto\{0,1\}$; 
demodulated bits $y^n$; optional information $\soft$. 
\STATE {\bf Output}: Decoding $c^{*,n}$.
\STATE $d\leftarrow 0$.
\WHILE{d=0}
\STATE $z^n\leftarrow$ next most likely binary noise effect sequence (which may depend on $\soft$)
 \IF{$\codebook(y^n\ominus z^n) = 1$}
\STATE  $c^{*,n}\leftarrow y^n\ominus z^n$ \& $d\leftarrow1$
\STATE{\bf return} $c^{*,n}$.
\ENDIF
\ENDWHILE
\STATE
\STATE
\hrule
\end{algorithmic}
\caption{Guessing Random Additive Noise Decoding. Inputs:
a demodulated channel output $y^n$; a code-book membership
function such that $\codebook(y^n)=1$ if and only if $y^n$ is in
the code-book; and optional statistical noise characteristics or soft
information, $\soft$. Output: decoded element $c^{*,n}$.}
\label{alg:pseudo-code}
\vspace{-0.5cm}
\end{figure}

Hard detection GRAND variants assume statistical knowledge of channel characteristics and recently variants have been established that leverage higher order modulation information to enhance their noise effect query order \cite{An22,Chatzigeorgiou22}. Consistent with conventional paradigms, all published soft detection variants of GRAND assume the availability of per-bit demapped soft information to inform the order of production of binary noise effect patterns. Using that information, Soft GRAND (SGRAND) \cite{solomon20SGRAND} produces optimally accurate decodings in an algorithm that is suitable for implementation in software, and so for code benchmarking purposes, but does not lend itself to implementation in circuits due to its need for dynamic memory. ORBGRAND \cite{duffy2021ordered,duffy2022ordered} is an approximation to SGRAND that is suitable for implementation in hardware by design and has already resulted in published circuits \cite{condo2021high,condo2022fixed,abbas22}. 
Here we show that ORBGRAND can be adapted to directly avail of soft information at the symbol level. In doing so, costly demapping is avoided.

\section{ORBGRAND for higher order modulations}
When working with binary soft information obtained from a receiver, ORBGRAND first determines the permutation that rank orders the $n$ hard demodulated bits in increasing reliability, which can be efficiently achieved with one of a broad range of algorithms, e.g. \cite{sorting14,Batcher1968}.
A piece-wise linear  model is then used to approximate the resulting reliability curve. The engineering merit of the approach is that within the context of the statistical model it is possible to efficiently sequentially construct putative noise effect patterns in decreasing order of likelihood \cite{duffy2022ordered}. These sequences can then be used within the GRAND framework, Fig. \ref{alg:pseudo-code}, to identify accurate decodings.

\begin{figure}[h]
    \centering
    \includegraphics[width= 0.4\textwidth]{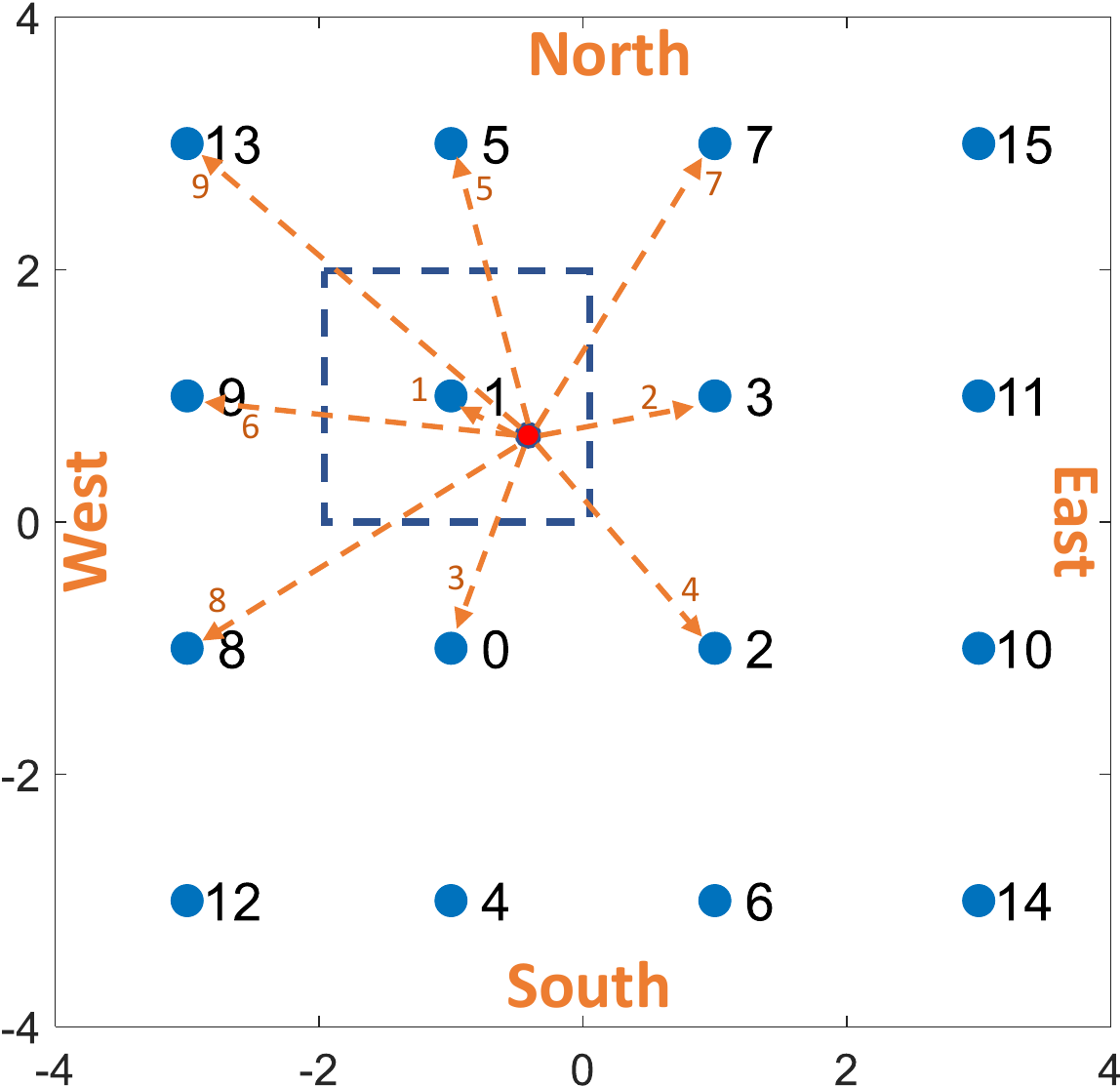}
    \caption{An illustration of 16 QAM, where the likelihood that a symbol was transmitted given the received signal
    is a monotonically decreasing function of their distance.}
    \label{fig:qam}
\end{figure}

When working with binary sequences, reliabilities are interpreted as relating to the likelihood that a bit is flipped, but they could, alternatively, be interpreted as the likelihood that the alternative possibility is correct. When working with higher order symbols, it is the latter interpretation that persists. That is, instead of searching for noise sequences by their likelihood, we construct symbol substitution sequences from reliability-ordered candidates. As in the hard-detection setting \cite{An22,Chatzigeorgiou22}, symbols neighbouring a received hard-detected signal are much more likely than those that are further away. As with soft detection binary decoding, the challenge is to efficiently and accurately incorporate symbol level soft information into a decoding process. 

In an additive complex white Gaussian noise (AWGN) channel with equally likely input sumbols, the probability of a candidate sequence $t^{n_s}$ is
\begin{align} \label{eq:orbSymProb}
    p_{X^{n_s}|Y^{n_s}}(t^{n_s}|Y^{n_s}) &= \frac{f_{Y^{n_s}|X^{n_s}}(Y^{n_s}|t^{n_s}) p_{X^{n_s}}(t^{n_s})}{f_{Y^{n_s}}(Y_{n_s})} \nonumber \\
    & \propto \prod_{i=1}^{n_s} e^{-|t_i-Y_i|^2/N_0}.
\end{align}
where $|Y_i-t_i|^2$ is the Euclidean distance between a candidate constellation symbol and the received signal in the complex plane, as illustrated in Fig. \ref{fig:qam}. 

The closest symbols to the received signals form the hard-detection sequence, which we denote $\Upsilon^{n_s}$. As a result of Eq. \eqref{eq:orbSymProb}, to rank order the likelihood of received sequences, it suffices to rank order by $\sum_{i=1}^{n_s}|t_i-Y_i|^2$ or, equivalently by
\begin{align} \label{eq:exceed}
\delta_i = |t_i-Y_i|^2 - |\Upsilon_i-Y_i|^2,
\end{align}
as the term $|\Upsilon_i-Y_i|^2$ is common to all sequences. We call $\delta_i$ the exceedance distance. Note that $\delta_i$ is positive owning to the fact that the hard-detected symbol $\Upsilon_i$ is defined to have the minimum distance to the received signal $Y_i$. Replacing any symbol in the hard-detected sequence $\Upsilon^{n_s}$ corresponds to generating a new candidate sequence leads to the increase of exceedance distance. Therefore, the likelihood of a candidate sequence is determined by summing up its exceedance distances, 
 $\sum_{i=1}^{n_s} \delta_i.$
One can verify that exceedance distance becomes log-likelihood reliability in BPSK modulation. 

For each received signal, its $\mu$ closest neighbours are considered for candidate symbol sequence construction. If $\mu=1$, then we are in the hard demodulated setting. Soft decoding occurs if $\mu > 1$, in which case the hard-detected symbol sequence is always the first to be checked for code-book membership. Other candidate sequences are constructed in the order of their exceedance distances values based on the statistical reliability model as in binary ORBGRAND. This results in a total of $n_s (\mu-1) = n (\mu-1)/m_s$ symbol level reliabilities that are rank ordered from least to greatest. If $\mu \leq m_s$, as will be seen to be sufficient to extract optimal decoding performance in practice, the collection of reliabilities to be sorted is shorter than in the binary case.

\begin{figure}[h]
    \centering
    \includegraphics[width= 0.4\textwidth]{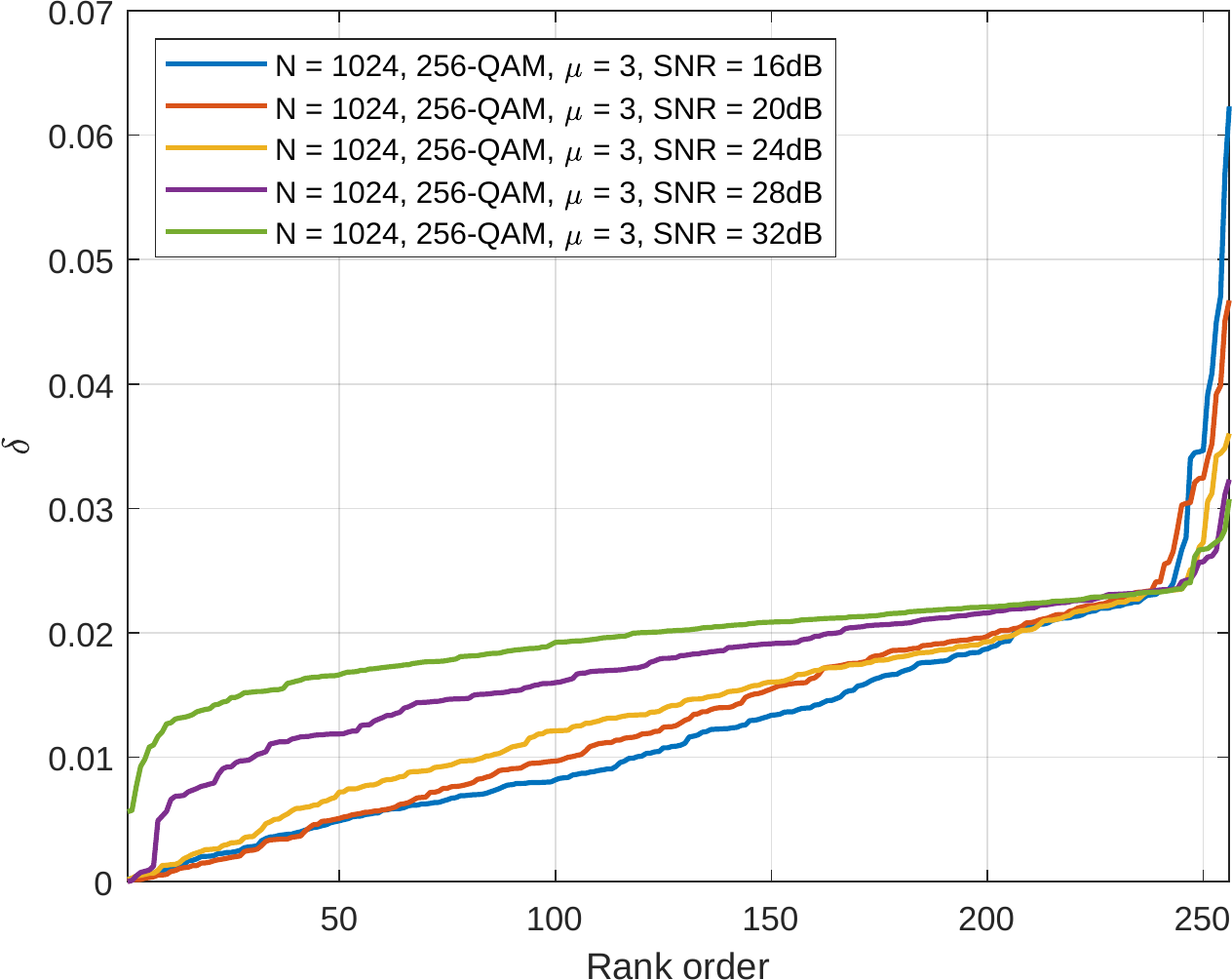}
    \caption{Rank ordered liabilities in term of exceedance distance for code-word length of 1024; With 256-QAM modulation and $\mu=3$, only 256 neighboring symbols are to be rank-ordered, as compared to 1024 soft bits to be rank-ordered after the demapper.}
    \label{fig:reliabilities}
\end{figure}

Fig. \ref{fig:reliabilities} plots empirical rank ordered symbol reliabilities in term of exceedance distance for 1024-bit code-word with 256-QAM modulation and $\mu=3$, as simulation in section \ref{sec:perf} will show that is sufficient for optimal decoding of [1024, 1002] codes. As in the binary case, a piece-wise linear statistical model provides a good description of the resulting data. Consequently, the original methodology for producing binary noise sequences of length $n$ that indicate which bits should be flipped \cite{duffy2022ordered} can be retained, but reinterpreted. With symbol level information, they instead produce binary sequences of length $n_s$ that indicate which symbols are to be to substitutes, that are interpreted as identifying symbols to be substituted for each noise-effect query. One issue arises in the higher order case of $\mu \geq 3$ that does not occur in the binary setting where, for example, two or more neighboring symbols of a hard-detected symbol are selected for its substitution. If ORBGRAND's pattern generator proposes substituting both simultaneously, that sequence is invalid and so skipped. Otherwise, the algorithm proceeds as before.

\section{Performance Evaluation}
\label{sec:perf}
For simulated performance evaluation, we consider 256-QAM modulation, corresponding to $m_s=8$ bits per symbol, and additive complex white Gaussian noise. For noise-effect pattern generation, we use 3-line ORBGRAND \cite{duffy2022ordered}. For error correcting codes, we use CA-Polar codes as they are the state-of-the-art short code with a dedicated binary soft detection decoder in CA-SCL ~\cite{KK-CA-Polar,TV-list,llr-ca-scl,leonardon2019fast, 6823099, 9186729, 8361464, 9020375}. We also consider codes for which there is no dedicated soft detection decoder exists, CRC codes and RLCs, demonstrating that they provide comparable error correction capability. 

\begin{figure}[h]
    \centering
    \includegraphics[width= 0.4\textwidth]{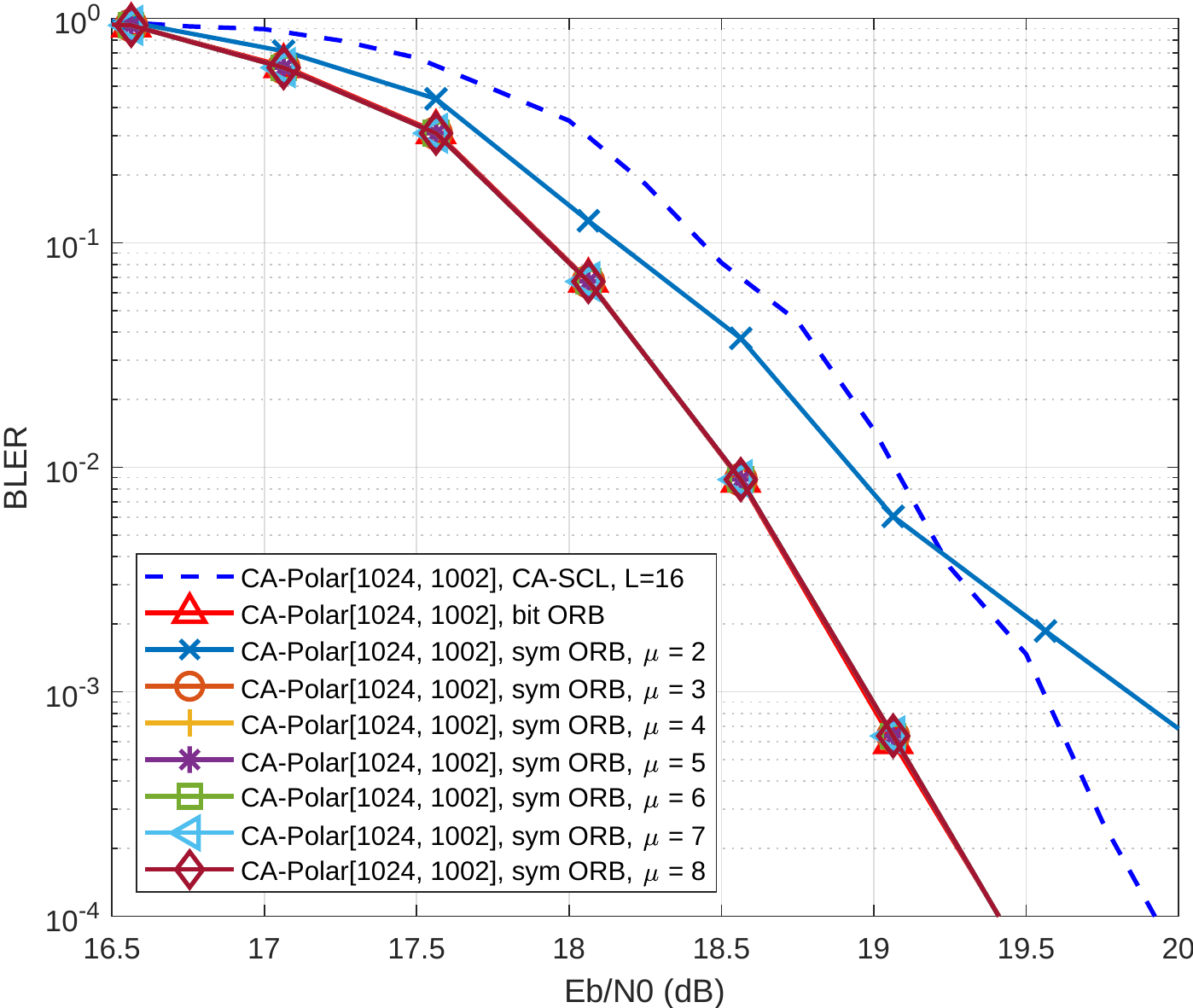}
    \caption{BLER for a CA-Polar[1024,1002] code over 256-QAM decoded with: CA-SCL with soft demapped bits; 3-line ORBGRAND with soft-demapped bits; 3-line ORBGRAND with symbol soft information and $\mu$ neighbours.}
    \label{fig:perf1024}
\end{figure}

Fig. \ref{fig:perf1024} presents block-error rate versus Eb/N0 for a CA-Polar[1024,1002] code. When operating with binary soft information ORBGRAND decoding outperforms CA-SCL as the latter does not fully avail of error correcting capability of the CRC aspect of the code, as previously reported \cite{duffy2022ordered}. By rank ordering the symbol-level reliabilities of as few as $\mu=3$ nearest neighbours per recevied symbol, i.e. a total of $n (\mu-1)/m_s = 256$ reliabilities rather than $n=1024$, ORBGRAND operating on symbols obtains the same BLER performance as ORBGRAND operating on soft demapped bits, but without the need for computationally involved demapping.

\begin{figure}[h]
    \centering
    \includegraphics[width= 0.4\textwidth]{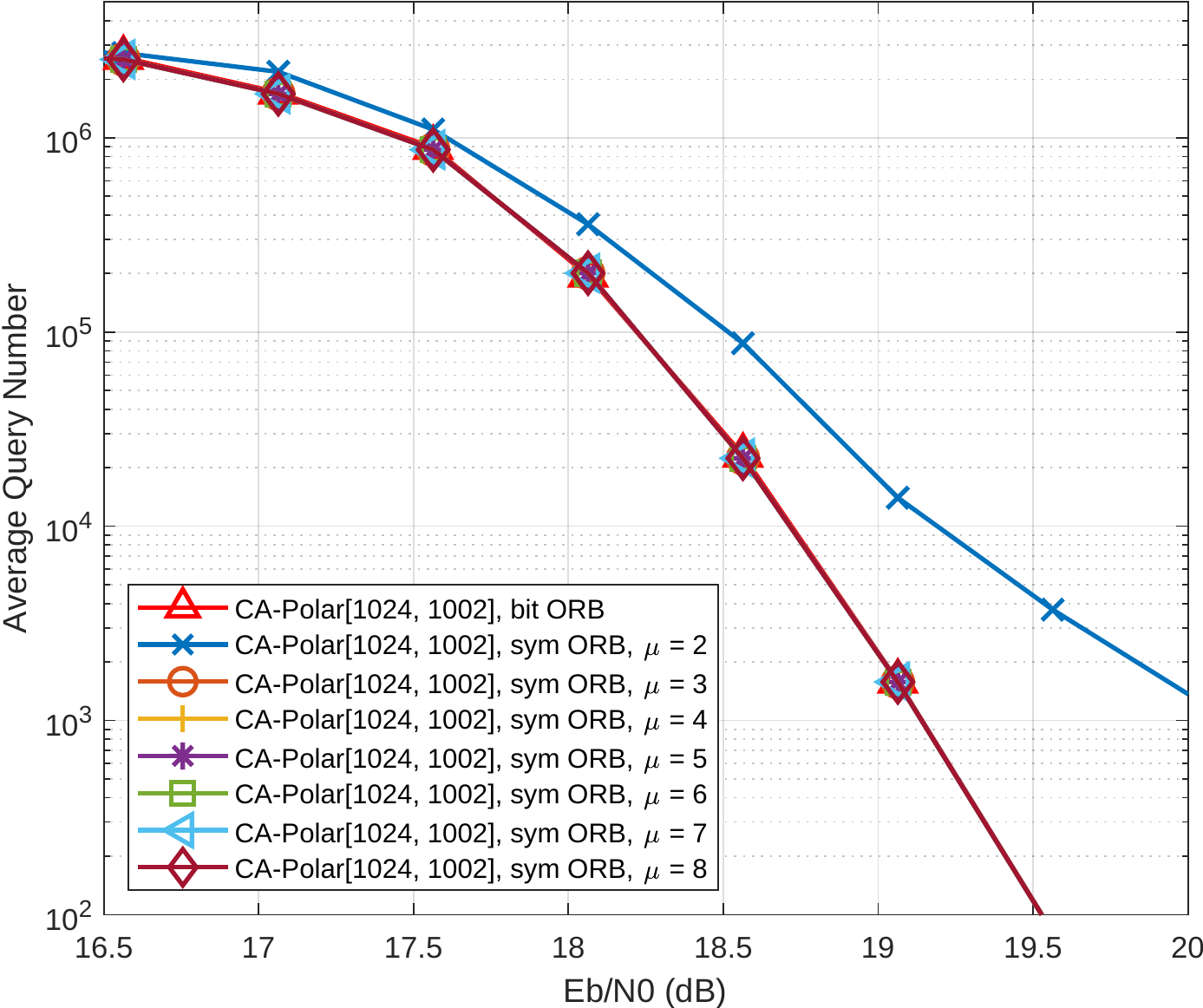}
    \caption{Complexity performance (average number of code-book queries to decoding) for a CA-Polar[1024,1002] code over 256-QAM decoded with 3-line ORBGRAND over symbol soft information and $\mu$ neighbours.}
    \label{fig:complex1024}
\end{figure}

As checking code-book membership is a computational simple operation for linear code-books, a standard metric for the complexity of GRAND algorithms is the number of code-book queries they make before identifying a decoding \cite{Duffy22,An22}. As SNR increases, the complexity of GRAND algorithms typically decreases precipitously, making it highly energy efficient in standard operating regimes \cite{Riaz21}. For the the setup used in Fig. \ref{fig:perf1024}, Fig. \ref{fig:complex1024} reports the associated complexity where it can be seen that the average number of code-book queries until a correct decoding is identified is nearly identical regardless of whether binary or symbol level information is used. At a standard operational BLER of $10^{-3}$, ORBGRAND makes approximately $3000$ queries, consistent with the GRAND complexity observations reported in \cite{An22, duffy2022ordered}.

\begin{figure}[h]
    \centering
    \includegraphics[width= 0.4\textwidth]{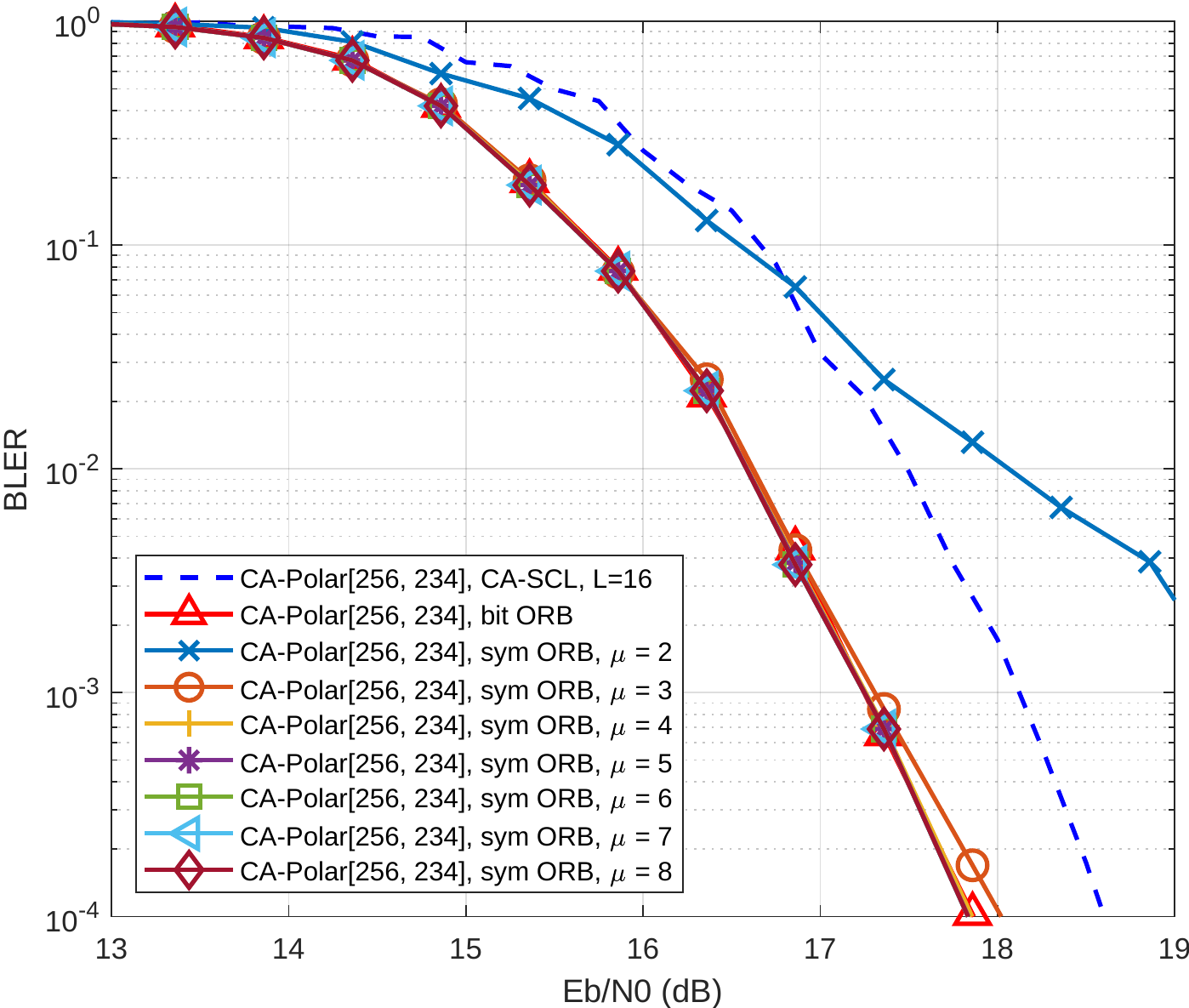}
    \caption{BLER for a CA-Polar[256,234] code over 256-QAM decoded with: CA-SCL with soft demapped bits; 3-line ORBGRAND with soft-demapped bits; 3-line ORBGRAND with symbol soft information and $\mu$ neighbours.}
    \label{fig:perf256}
\end{figure}

\begin{figure}[h]
    \centering
    \includegraphics[width= 0.4\textwidth]{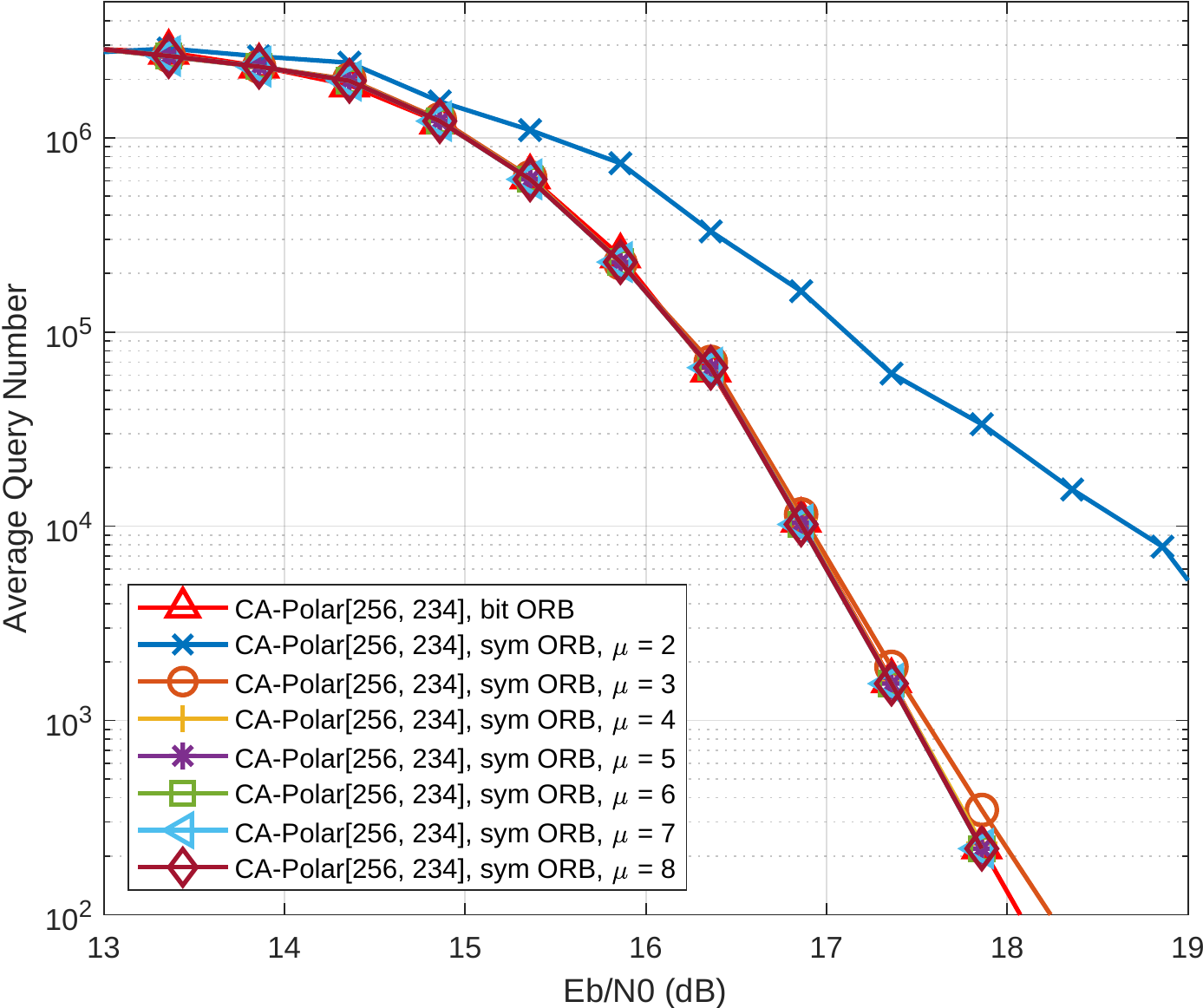}
    \caption{Complexity performance (average number of code-book queries to decoding) of a CA-Polar[256,234] code over 256-QAM decoded with 3-line ORBGRAND with symbol soft information and $\mu$ neighbours.}
    \label{fig:complex256}
\end{figure}

Fig. \ref{fig:perf256} and Fig. \ref{fig:complex256} report equivalent results, but for a shorter, lower-rate CA-Polar[256,234] code. Here, with as few as $\mu=4$ nearest neighbours per received signal, ORBGRAND with symbol level soft information performs as well as its binary counterpart. In the symbol level version, this necessitates that $96$ exceedance distances  be rank ordered rather than $256$ reliabilities as in the binary setting, yet performance is ultimately the same. This comes about as by operating on symbols, the decoupling of bits that occurs with demapping is avoided.

\begin{figure}[h]
    \centering
    \includegraphics[width= 0.4\textwidth]{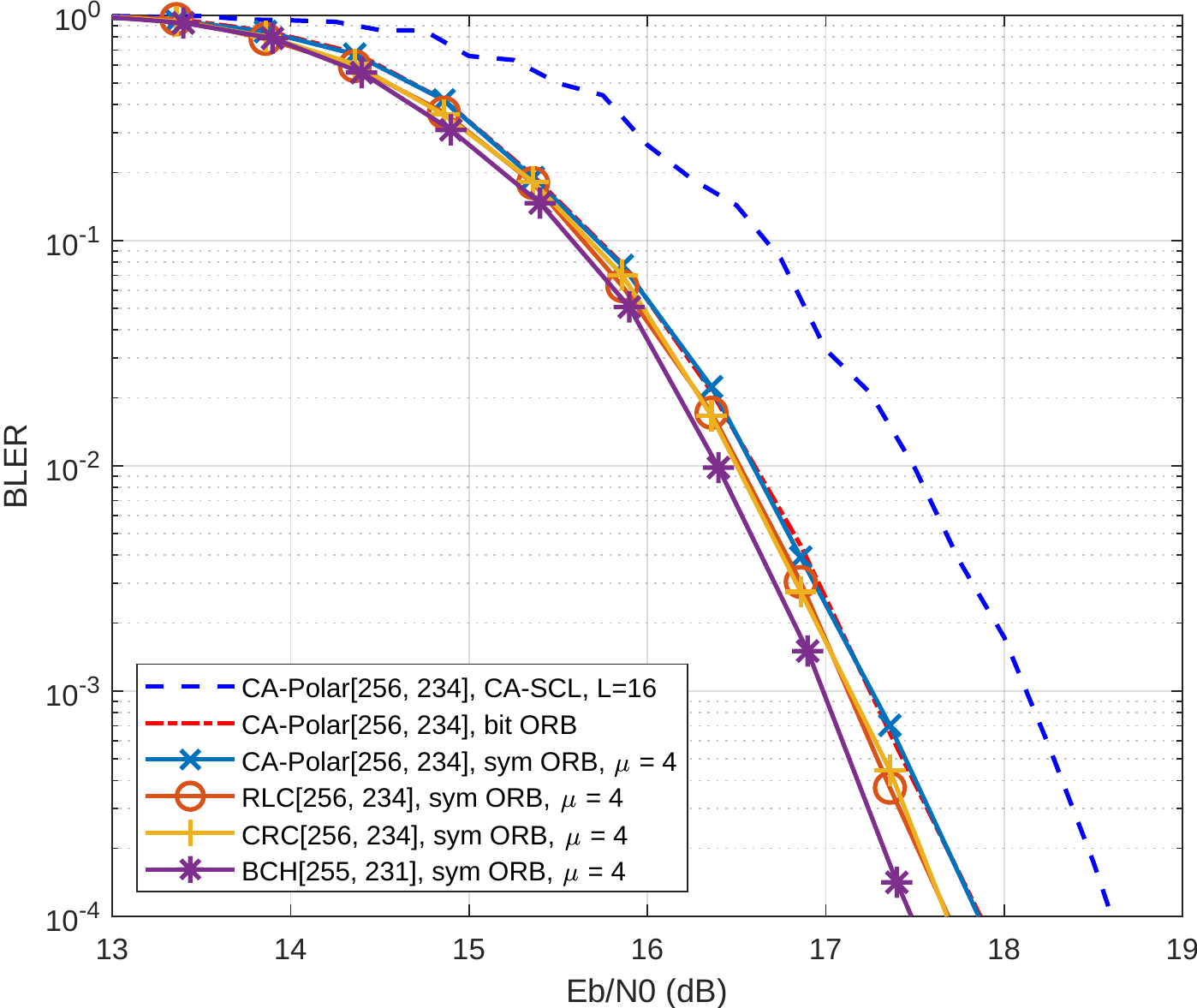}
    \caption{BLER for CA-Polar, RLC and BCH codes: CA-SCL with soft demapped bits; 3-line ORBGRAND with soft-demapped bits for CA-Polar[256,234]; 3-line ORBGRAND with symbol soft information and $\mu=4$ neighbours for CA-Polar[256,234], RLC[256,234], CRC[256,234] and BCH[255,231].}
    \label{fig:perf256other}
\end{figure}

Modern applications, including augmented and virtual reality,
vehicle-to-vehicle communications, the Internet of Things, and
machine-type communications, have driven demand for Ultra-Reliable
Low-Latency Communication (URLLC)
\cite{durisi2016toward,she2017radio,chen2018ultra,parvez2018survey,medard20205}.
To enable these technologies requires the use of shorter, higher-rate codes.
This has placed renewed focus on conventional codes \cite{short_fec, bch_m2m} 
as well as CA-Polar codes. Results from GRAND algorithms
with binary soft information have found that most code-structures provide
similar block error performance, \cite{an21, Papadopoulou21}. This finding translates to the setting
where soft symbol level reliability information is used in the decoding,
as demonstrated by the 
results for an RLC and a CRC shown in Fig. \ref{fig:perf256other}, which corresponds to
the CA-Polar setting in Fig. \ref{fig:perf256}. A BCH is also presented with a slightly different setting due to its code rate limitation. This suggests that a much larger palette of
codes is available through which URLLC can be delivered.

\section{Discussion}
\label{sec:disc}
Existing soft detection decoders rely on receivers providing per-bit reliability information
as evaluated through a computationally costly process that must be performed for each received
symbol. Here we establish that soft decoding without soft demapping is possible with ORBGRAND. 
By retaining symbol level information, a shorter list of reliabilities is needed to provide identical 
decoding performance. Moreover, as the approach is suitable for decoding any moderate redundancy code
and similar performance is extracted from a broad range of code-structures, this offers one possible approach
in the delivery of URLLC.

\bibliographystyle{IEEEtran}
\bibliography{ORBGRAND,grand}

\end{document}